\def \NH {N_{\rm H}}

\documentstyle[12pt,aasms4]{article}

\tighten
\eqsecnum

\begin{document}

\title{OSSE OBSERVATIONS OF THE ULTRALUMINOUS 
\\INFRARED GALAXIES ARP 220, MRK 273, AND MRK 231}

\author{C. D. Dermer\altaffilmark{1}, J. Bland-Hawthorn\altaffilmark{2}, J.
Chiang\altaffilmark{1,3}, \& K. McNaron-Brown\altaffilmark{1,4}}

\altaffiltext{1}{E. O. Hulburt Center for Space Research, Code 7653, Naval Research
Laboratory, Washington, DC 20375-5352 USA}
\altaffiltext{2}{ Anglo-Australian Observatory, P. O. Box 296, Epping, NSW 2121, Australia}
\altaffiltext{3}{NRL/NRC Research Associate}
\altaffiltext{4}{CSI, George Mason University, Fairfax, VA  22030 USA}

\begin{abstract}

We report results of soft gamma-ray observations of the ultraluminous infrared galaxies Arp 
220, Mrk 273, and Mrk 231 in order to test whether
the infrared radiation from these sources originates from buried active galactic nuclei 
(AGNs).
Only upper limits are measured, implying that the emergent soft gamma-ray luminosities are 
1-2
orders of magnitude smaller than the infrared luminosities. Monte Carlo simulations of 
radiation
transport through tori are used to infer the minimum column densities $\NH$ required to 
block
transmission of soft gamma-rays from a buried AGN, assuming that spectra of AGNs in such 
sources
are similar to those of radio-quiet quasars. Lack of measured gamma-ray emission provides 
no
supporting evidence for the existence of buried AGNs in these galaxies, but is
consistent with an  origin of the infrared luminosity from starburst activity.
 
\end{abstract}


\section{Introduction}

The enormous infrared luminosities ($L_{\rm IR}  \gg 10 ^{10} L_\odot)$  discovered
in  merging galaxies could originate from enhanced fueling of supermassive
black holes surrounded by large column densities of dust and gas, dust heated by starburst
activity, or galactic winds driven by starburst activity or the kinetic energy of colliding 
galaxies
(e.g., Soifer et al. \markcite{Soife84}1984, Sanders et al. \markcite{Sande88a}1988a).  
Detection of
time-variable high-energy radiation from ultraluminous infrared ($L_{\rm IR} > 10^{12} 
L_\odot$)
galaxies, or ULIGs, would prove the existence of AGNs in these systems and support the 
interpretation
that quasar activity is related to mergers (Sanders et al. \markcite{Sande89}1989).  If 
vigorous star
formation or galactic winds  rather than nuclear activity produce the infrared  emission, 
significantly
less hard X-ray and soft gamma-ray emission is expected (Rephaeli et al. 
\markcite{Repha91}1991;
Rephaeli, Ulmer, \& Gruber
\markcite{Repha94}1994). Photons with energies $\lesssim 10$ keV are attenuated by  
photoelectric
absorption in Solar composition material with  hydrogen column densities $\NH \gtrsim 
10^{23}$
cm$^{-2}$.  Compton scattering opacity dominates photoelectric absorption at photon 
energies $\gtrsim
10$ keV, but the Klein-Nishina decline in the Compton cross section above $\sim 100$ keV 
makes the
escape of higher-energy photons more probable. Gamma-ray observations therefore provide the 
best means
for detecting a dust-enshrouded AGN (see, e.g., Krolik, Madau, \& Zycki
\markcite{Kroli94}1994). 

Here we report observations of Arp 220,  Mrk 273, and Mrk 231 using the Oriented 
Scintillation Spectrometer Experiment (OSSE) on the {\it Compton Gamma Ray
Observatory} ({\it CGRO}).  These three ULIGs are among the most luminous galaxies
within 200 Mpc at any wavelength, with bolometric
luminosities  almost two orders of magnitude higher than ordinary spiral galaxies and 
similar to
that of quasars. All three galaxies show signs of interactions and mergers, for example, a
double nucleus in Arp 220, a jetlike protusion in Mrk 273 probably due to a disk distorted 
by
tidal effects, and tidal tails in Mrk 231.  Large quantities of molecular gas, which are
necessary to fuel AGNs or produce vigorous starburst activity, are found in all three
systems.  Signatures of Seyfert-like
nuclei are found in Mrk 273 and Mrk 231, with the latter source displaying H$\alpha$ line 
widths
characteristic of a Seyfert 1 nucleus (for a recent review, see Sanders \& Mirabel
\markcite{Sande97}1997).  

Table 1 gives the infrared luminosities, redshifts, and distances of Arp 220, Mrk 273, and 
Mrk 231,
along with the measured upper limits for the  ratios of the 50-200 keV gamma ray 
luminosities to the
8-1000 $\mu$m infrared luminosities.  Our non-detections place limits on the luminosity of 
the central
source and the column density of intervening gas.  We describe the observations in \S 2, 
and in
\S 3 we construct multiwavelength spectral energy distributions of these galaxies.  Our
Monte Carlo simulation of a nuclear source surrounded by a gaseous torus is described in \S 
4,
and the implications of the observations are discussed in \S 5.

\section{Observations}

OSSE, one of four instruments on {\em CGRO}, is designed to detect gamma rays in the $0.05 
- 10$~MeV
range.  OSSE comprises four independent phoswich spectrometers of identical design that are 
each
actively shielded and passively collimated. Tungsten collimators define a 
$3.8^{\circ}\times
11.4^{\circ}$ full-width at half-maximum gamma-ray aperture. Table 2 summarizes the OSSE
observations, plotted in Fig. 1, of the three ULIGs observed to date.  The
upper limits for the fluxes in the 50 -- 100 keV and 100 -- 200 keV energy ranges are given 
 at the 95\%
(2$\sigma$) confidence level, assuming an intrinsic source spectrum with photon spectral 
index $\alpha =
2$.  The upper limits are constant within $\approx 20$\% for $1\lesssim \alpha \lesssim 3$. 
 Upper
limits in the 200 -- 400 keV and 400 -- 700 keV ranges are also shown in Fig. 1, but are 
much less
constraining than the lower energy data for typical AGN X-$\gamma$ ray spectra with $\alpha 
\approx 2$.

The spectra in the sample were analysed uniformly, with similar
data-selection criteria applied to all observations according to the procedures
described by Johnson et al. (\markcite{Johns93}1993).
Quadratic interpolation in time among the measured background intervals 
is used to estimate the background during the source observation. Standard background 
offsets of
4.5$^\circ$ on either side of the source positions along the detector scan plane were used. 
The
background estimates are then subtracted from the associated source accumulations to form
two-minute difference  spectra.  These spectra are further screened for environmental 
effects and transient phenomena.  Screened 2-minute spectra are then summed 
into daily average spectra and finally into spectra averaged over the entire
observation interval or, as in the case of Arp 220, the entire set of observations.

\section{Multiwavelength Spectra}
 
Figure 1 shows multiwavelength $\nu L_\nu = 4\pi d_L^2 \nu F_\nu$ spectra of the three
ULIGs observed with OSSE, including the 2$\sigma$ upper limits derived in the present
analysis.  Values of $\nu L_\nu$ are derived assuming isotropic source emission with a 
luminosity
distance $d_L = 2c[z+1-(z+1)^{1/2}]/H_0$, appropriate to a critical density cosmology with 
zero
cosmological constant.  We use a Hubble constant $H_0 = 75$ km s$^{-1}$ Mpc$^{-1}$.  The
radio, 1250 $\mu$m, submillimeter, and infrared and optical data are from Condon et al.
(\markcite{Condo91}1991), Carico et al. (\markcite{Caric99}1988), Rigopoulou et al.
(\markcite{Rigop96a}1996a), and  Sanders et al. (\markcite{Sande88a}1988a), respectively. 
The 0.1-4.5 keV
{\it Einstein} data for Arp 220 are from Eales
\& Arnaud (\markcite{Eales88}1988), and the {\it ROSAT} data for Mrk 273 in the range 0.1 - 
2.0 keV and
for Mrk 231 in the range 0.1-2.4 keV are from Turner, Urry, \& Mushotzky 
(\markcite{Turne93}1993) and
Rigopoulou et al. (\markcite{Rigop96a}1996a), respectively. {\it HEAO} A-1 upper limits in 
the 2-10 keV
range are from Rieke (\markcite{Rieke88}1988), and additional data for Arp 220 are from 
Rigopoulou et al.
(\markcite{Rigop96a}1996a), who also list the beam sizes for the different detectors.

As can be seen from Figure 1, ULIGs show an extraordinary feature in their $\nu
L_\nu$ spectra peaking near 100 $\mu$m. The upper limits to the
gamma-ray luminosities are $\sim 1-2$ orders of magnitude less than the infrared 
luminosities $L_{\rm IR}$, but are also $\sim 2$ orders of magnitude greater than the soft 
X-ray
luminosities. The $\sim 10^{17}$ Hz X-rays are probably from hot gas driven by starburst 
activity or
from AGN emission scattered by high latitude gas. A direct nuclear origin would require a 
low column
density, implying an AGN luminosity orders of magnitude less than the IR luminosity, in 
which case a
dust-enshrouded AGN could not be the primary IR power source.  

\section{Monte Carlo Simulation of Photon Transport}

We test whether the IR luminosity in these sources originates from an AGN surrounded by 
large
columns of gas and dust. The Monte Carlo model simulates a central source of continuum 
radiation
surrounded by a uniform torus with circular cross-section. The spectrum of the central
source is represented by a two-component accretion-disk spectrum of the form
$$\dot N(E) = k_1 E^{-2/3}\exp(-E/E_1) + k_2  E^{-\alpha} \exp(-E/E_2)H[E-E_1]\;,\eqno(1)$$

\noindent where $H[E-E_1] = 1$ if $E>E_1$ and $H[E-E_1] = 0$ otherwise. The first term on
the rhs of equation (1), which we denote $\dot N_1(E)$, represents the cool, optically thin
Shakura-Sunyaev spectrum thought to produce the enhanced UV (``big blue bump") emission 
observed in
Seyfert galaxies and quasars. The second term, which we denote
$\dot N_2(E)$, represents the X-ray and gamma-ray emission observed from such sources. The 
normalization
of the components is given by the condition
$$L_i = \int_0^\infty dE \cdot E \cdot \dot N_i(E)\;,\; i = 1,2\;.\eqno(2)$$

\noindent Defining the total
luminosity $L_{\rm tot} = L_1 + L_2$ and $f = L_1/L_2$, we find that the unabsorbed X-ray 
and gamma-ray
$\nu L_\nu$ spectral component is given to good approximation by
$$(\nu L_\nu)_{X\gamma} \cong {L_{\rm tot}\over
(1+f)[\Gamma(2-\alpha)-{u^{2-\alpha}\over 2-\alpha}+{u^{3-\alpha}\over 
3-\alpha}]}\;({E\over
E_1})^{2-\alpha}\exp(-E/E_2)\;,\eqno(3)$$

\noindent provided that $\alpha < 2$ and $u \equiv E_1/E_2 \ll 1$. 

We choose $E_1 = 30$ eV for the cool outer blackbody spectrum, $\alpha = 1.9$ for the 
photon spectral
index of the X-$\gamma$ component, and consider two values for
$E_2$ based on OSSE observations of Seyfert 1 galaxies (Gondek et al. 
\markcite{Gonde96}1996), namely
$E_2 = 100$ keV and 400 keV. We assume solar abundances for the neutral torus material 
using the
photoelectric absorption cross-sections of Morrison \& McCammon (\markcite{Morri83}1983) 
and the full
Klein-Nishina cross section (see Chiang, Dermer, \& Skibo \markcite{Chian97}1997 for a full 
discussion of
the model).  Our conclusions are only weakly sensitive to $\alpha$ between 1.5 and 2, to 
$E_2$ in the
range 50 keV -1 MeV and to the value of $E_1$, provided that $\alpha < 2$. 

The emergent spectra are generated from escaping photons which have directions within 
$5^\circ .7$ of the
equatorial plane of the torus. Results are plotted in Figure 2 in the form $\nu L_\nu 
(1+f)/L_{\rm tot}$
suggested by equation (3). The torus opening angle is 45$^\circ$ in Figure 2; results for 
torus opening
angles between $\approx 0^\circ$ and 60$^\circ$ degrees differ at most by a factor of two 
at $\NH =
10^{25}$ cm$^{-2}$, and by less at $\NH \lesssim
10^{25}$ cm$^{-2}$.  For comparison with data, we identify $L_{\rm tot}$
with $L_{\rm IR}$ given in Table 1; in other words, we test the hypothesis that the IR 
luminosity
originates from a buried AGN. We let
$f =9$, based on the results of Elvis et al. (\markcite{Elvis94}1994) for the mean 
radio-quiet quasar and
Seyfert energy distributions.\footnote[1]{\rm The average and standard deviation of the 
ratios of the
0.1-1
$\mu$m and 1-10 keV luminosities for the 19 radio-quiet and Seyfert galaxies in the Elvis 
et al. sample
are 7.4 and 3.8, respectively.  The Elvis et al. results imply that $\gtrsim 95$\% of such 
sources have
$f
\lesssim 15$, although the bolometric correction adds additional uncertainty. }  Thus 10\% 
of the
total AGN luminosity is emitted in the X-$\gamma$ component. 

The histograms in Fig. 1 show the $\NH = 10^{24}$ cm$^{-2}$ simulations overlaid on the 
$\nu L_\nu$
spectra of the ULIGs observed with OSSE for $E_2 = 100$ and 400 keV.  Column densities $\NH 
>
10^{23.5}$-$10^{24}$ cm$^{-2}$ are necessary to agree with the hard X-ray upper limits 
(Rieke
\markcite{Rieke88}1988) for our standard spectrum.  Strong constraints on the nature of the
central source and obscuring column are provided by soft gamma-ray observations of Arp 220. 
 Either the
column density $\gtrsim 10^{25}$ cm$^{-2}$ or, if $\NH \lesssim 10^{24}$ cm$^{-2}$, then 
$\gtrsim$ 80\%
of the total $L_{\rm IR}$ from Arp 220 is produced by non-AGN activity.  The constraints on 
AGN
activity in Mrk 273 and Mrk 231 are weaker than for Arp 220.

\section{Discussion and Summary}

Determining the source of the far-infrared radiation in ULIGs is important for questions of 
galaxy
and quasar evolution. Ultraluminous IRAS galaxies seem to provide the clearest 
observational link between
galaxy mergers and nuclear activity (Stockton \markcite{Stock90}1990). It has long been 
known that
galaxies with active nuclei have  large infrared luminosities (Rieke \& Low
\markcite{Rieke72}1972), and it has been argued that  ULIGs
are the parent population of quasars (Soifer et al. \markcite{Soife86}1986; Sanders et al.
\markcite{Sande89}1989). The coincidence between the
bolometric luminosities of ULIGs and the  luminosities of bright Seyfert galaxies
(Sanders et al. \markcite{Sande88b}1988b) suggests that the infrared emission from these 
galaxies is
reprocessed UV and X-radiation emitted by active galactic nuclei. 

One possible explanation for the non-detection of gamma-rays from Arp 220, Mrk
273, and Mrk 231 is that the AGNs are hidden behind a very large column of gas with $\NH 
\gtrsim
10^{24}$ cm$^{-2}$. The total mass of gas is limited, however, by observations at other 
wavelengths.
Dynamical masses inferred from observations of Arp 220 require that the gas mass not
exceed $ \sim 2 \times 10^9$ M$_{\odot}$ within 350 pc (Shier, Rieke, \& Rieke 
\markcite{Shier94}1994),
implying that $\NH\lesssim 3\times 10^{23}$ cm$^{-2}$ for a covering factor of one. 
Observations of
millimeter wave CO emission suggest that the gas mass in Mrk 273 is not more than 
$10^{10}$ M$_{\odot}$ for Mrk 273 (Rigopoulou  et al. \markcite{Rigop96b}1996b). As the gas 
distributions
are typically a few hundred parsec in extent (Scoville  et al. \markcite{Scovi91}1991), the 
implied
column density is $\lesssim 2\times 10^{24}$  cm$^{-2}$, again assuming uniform covering.  
In the case
of Mrk 231,  a gas mass $\approx 3.4\times 10^9$ M$_\odot$ is measured within a sphere of 
radius 420
pc (Byrant \& Scoville \markcite{Bryan96}1996).  The molecular gas must be
distributed in a disk with additional scattering from high latitude gas to account for the
spectropolarimetry observations of Smith et al. (\markcite{Smith95}1995) and the 2 
magnitudes of
visual extinction inferred from optical reddening of continuum and line emission 
(Boksenberg
et al. \markcite{Bokse77}1977). Rudy et al. (\markcite{Rudy85}1985) find that $\NH \sim 
10^{22}$
cm$^{-2}$ from spectroscopy of Na in Mrk 231, which is consistent with measured 
absorption-line features
characteristic of broad absorption-line QSOs (Smith et al. \markcite{Smith95}1995), which 
are found to
have column densities $\sim 10^{23}$ cm$^{-2}$ (e.g., Green
\& Mathur
\markcite{Green96}1996) .  

The covering factor for the gas cannot, in a statistical sense, be much smaller than 1, due 
to the
presence of more IR-luminous galaxies than optically selected AGN in the local universe  at 
a bolometric
luminosity of $10^{12}$ L$_{\odot}$ (Schmidt \& Green \markcite{Schmi83}1983; Soifer  et 
al.
\markcite{Soife87}1987). We cannot rule out the possibility that for the
specific cases of Arp 220 and Mrk 273, patchy column densities have hidden the nuclear 
sources of the
IR radiation. The column density to the
nucleus of Mrk 231 is $< 10^{23}$ cm$^{-2}$, so that the hard X-ray observations reported 
by
Rieke (\markcite{Rieke88}1988) constrain the luminosity of any buried AGN to contribute 
$\lesssim 15$\%
of the IR power.  

To summarize, we have placed limits on the gamma-ray fluxes of the 
ultraluminous infrared galaxies Arp 220, Mrk 273, and Mrk 231, with 2$\sigma$ upper limits 
in the 50-200
keV band of $3\times 10^{43}$, $1.6\times 10^{44}$, and $1\times 10^{45}$ ergs s$^{-1}$, 
respectively.
The measured gamma-ray to infrared luminosity ratios do not provide evidence in favor of 
the
interpretation that the ultraluminous infrared radiation is reprocessed AGN emission.  We 
are unable to
rule out the possibility that Arp 220, Mrk 273, and Mrk 231 are quite variable in 
gamma-rays, and that
our failure to detect them is due to observing them in the low state. Additional {\it CGRO} 
OSSE
observations would constrain the duty cycle of putative AGN activity.  Other possibilities 
to account
for our nondetections are that patchy columns obscure our lines-of-sight to the nuclear 
sources in Arp
220 and Mrk 273, or that accretion-disk spectra in ULIGs are unlike
other AGNs. By penetrating to column densities corresponding to $A_V \sim 5000$,
the OSSE results strengthen recent {\it ISO} observations (Lutz et al. 
\markcite{Lutz96}1996; Sturm et
al. \markcite{Sturm96}1996) which indicate that most ULIGs, including Arp 220, are powered 
by
starbursts.  We conclude that there is no evidence from the hard X-ray and
$\gamma$-ray observations that AGNs provide the source of the
far-infrared luminosity in Arp 220, Mrk 273, and Mrk 231.

\acknowledgments   We thank Sylvain Veilleux, Lisa Shier, and the anonymous referee for 
discussions and
comments, and Jeff Skibo for allowing us to modify his Monte Carlo simulation code.  The 
work of C. D.
was supported by the Office of Naval Research and NASA {\it Compton Observatory} Guest 
Investigator
grants DPR S-57770-F, DPR S-67006-F, S-30928-F and S-14643-F. J. B. H. acknowledges a 
Fullam Award from
the Dudley Observatory.

\clearpage

\begin{table*}[tbh]
\begin{center}
\caption{Ultraluminous Infrared Galaxies Observed with OSSE}
\vskip0.3in

\begin{tabular}{||l|c|c|c|c||} \hline

 & log [${L_{\rm IR}\over L_\odot}$]$\tablenotemark{a}$ &  z  & d (Mpc)$\tablenotemark{b}$ 
&
$({L_{\gamma}\over L_{\rm IR}})_{\rm max}\tablenotemark{c}$  
\\   \hline
 &                      &     &   &   \\   
Arp 220	& 12.19 & 0.0181 & 73 & 0.005\\
 &                      &     &   &   \\
Mrk 273 & 12.14 & 0.0376 & 151 & 0.03 \\ 
 &                      &     &   &   \\
Mrk 231 & 12.52 & 0.0410 & 166 & 0.08\\ \hline
\end{tabular}
\end{center}
\tablenotetext{a}{~ $L_{\rm IR}$ is infrared luminosity between 8 and $1000~\mu{\rm}$m from 
Sanders
et al. (\markcite{Sande88a}1988a).}
\tablenotetext{b}{~$H_0 = 75$ km s$^{-1}$ Mpc$^{-1}$.}
\tablenotetext{c}{~$L_{\gamma}$ represents 2$\sigma$ upper limits to the gamma-ray 
luminosity in
the range 50-200 keV.}
\end{table*}

\clearpage

\begin{table*}[tbh]
\begin{center}
\small
\caption{Observing Log and Gamma-Ray Flux Upper Limits of Ultraluminous Infrared Galaxies}
\begin{tabular} {lcllccc}
\\
& Viewing & & & Live Time & \multicolumn{2}{c}{Photon Flux$\tablenotemark{b}$} \\
Source & Period & \multicolumn{2}{c}{UT Date Interval} & on source$\tablenotemark{a}$ &
 ($0.05 - 0.10$ MeV) &  ($0.10 - 0.20$ MeV) \\
\hline
\hline
Arp 220 &  &  &  & 6.96 & $< 1.2$ & $< 0.7$  \\
     & 519 & 1996 Apr 23 & 1996 May 07 &   2.73 & $ <  1.9$ & $ <  1.2$ \\
& 531 & 1996 Oct 03 & 1996 Oct 15 &   2.35 & $ <  2.1$ & $ <  1.2$ \\
& 604 & 1996 Dec 05 & 1996 Dec 10 &   0.47 & $ <  4.8$ & $ <  2.9$ \\
& 605 & 1996 Nov 26 & 1996 Dec 03 &   1.41 & $ <  2.9$ & $ <  1.8$ \\
\hline
     Mrk 231 & 604 & 1996 Dec 05 & 1996 Dec 10 &   0.82 & $ <  8.4$ & $ <  5.3$ \\
\hline
     Mrk 273 & 515 & 1996 Feb 20 & 1996 Mar 05 &   4.87 & $ <  1.5$ & $ <  0.9$ \\
\hline
\\
\end{tabular}
\end{center}
\tablenotetext{a}{~In units of $10^5$ detector-seconds.}
\tablenotetext{b}{~In units of $10^{-4} ~\mbox{photons cm}^{-2}~
      \mbox{s}^{-1}$; upper limits are $2\sigma$.}

\end{table*}

\eject

\noindent \bf{Figure~Captions}

\figcaption{Multiwavelength $\nu L_\nu$ spectra of Arp 220, Mrk 273, and Mrk 231. The Monte 
Carlo
simulations show the fluxes expected if AGNs with X-ray/gamma-ray luminosities equal to 
10\% of
the measured 8-1000 $\mu$m IR luminosity are buried behind a torus with a neutral hydrogen 
column
density of $ 10^{24}$ cm$^{-2}$. The solid and dotted curves refer to source spectra with 
100 and 400 keV
cutoffs, respectively.  See text for references to data and details of the model.}

\figcaption{Monte Carlo simulations of photon transport from a central source located at 
the
center of a torus with a 45$^\circ$ opening angle.  The curves are labeled by values of the
neutral hydrogen column density along a line passing radially through the torus center, and 
the dark and
light histograms represent source spectra with 100 and 400 keV exponential cutoffs, 
respectively.}

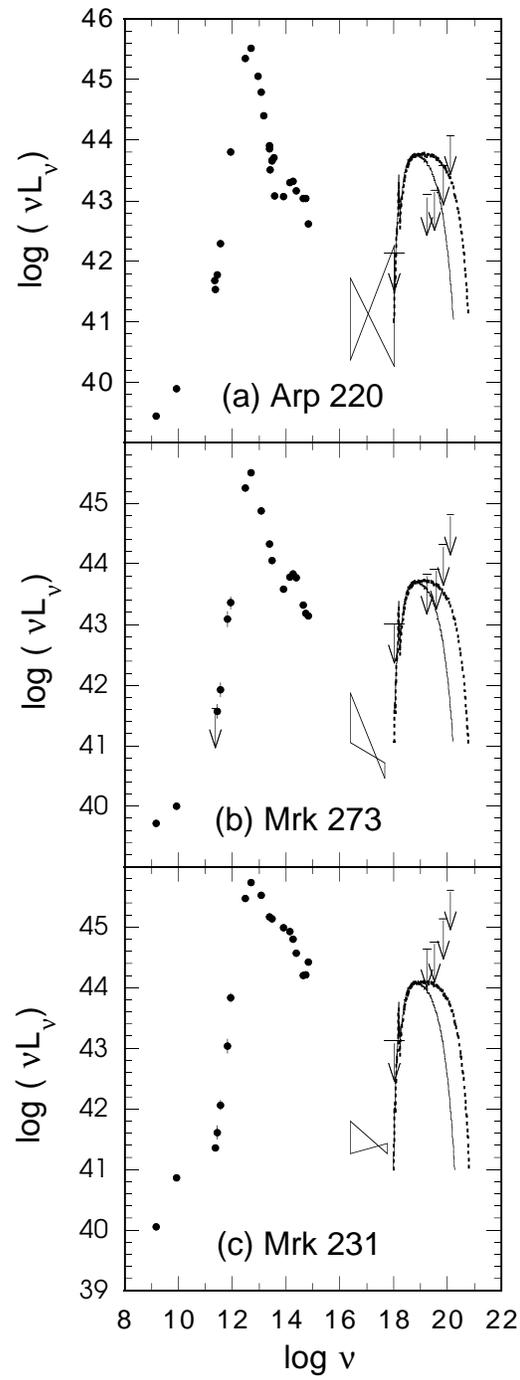
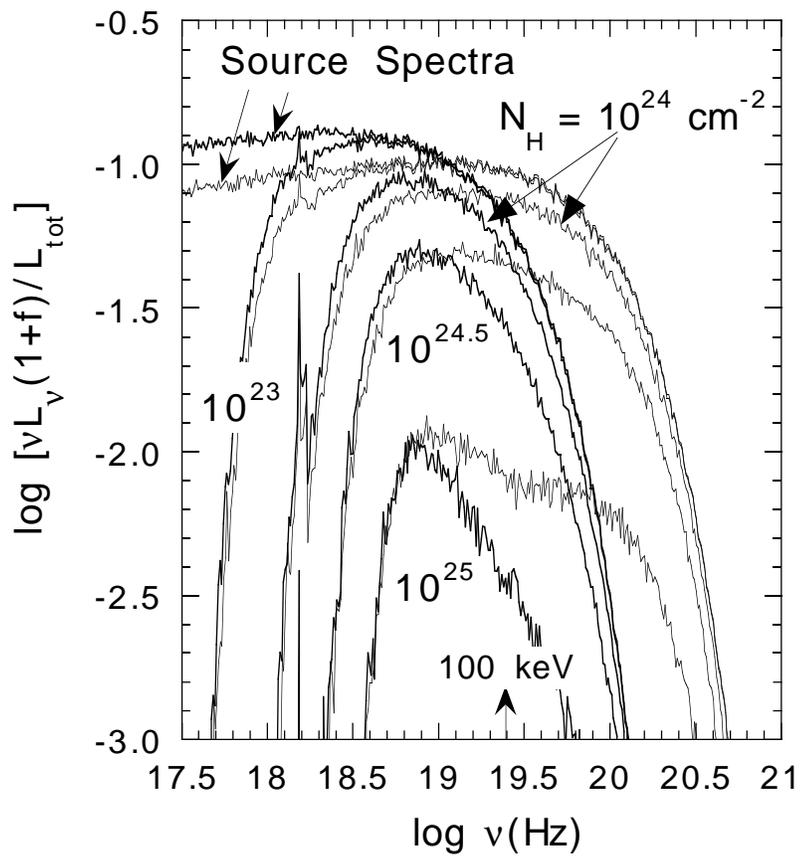
\end{document}